# Real-time monitoring of stress evolution during thin film growth by in situ substrate curvature measurement

Elisa Gilardi[1], Aline Fluri[1], Thomas Lippert[1,3,4], Daniele Pergolesi[1,2]*

[1] Research with Neutrons and Muons Division, Paul Scherrer Institut, 5232 Villigen PSI, Switzerland

[2] Energy and Environment Research Division, Paul Scherrer Institut, 5232 Villigen PSI, Switzerland

[3] Department of Chemistry and Applied Biosciences, Laboratory of Inorganic Chemistry, ETH Zürich, Vladimir-Prelog-Weg 1-5/10, 8093 Zürich, Switzerland

[4] International Institute for Carbon Neutral Energy Research (WPI-I2CNER), Kyushu University, 744 Motooka, Nishi-ku, Fukuoka 819-0395, Japan

*daniele.pergolesi@psi.ch

**Abstract**

Strain engineering is the art of inducing controlled lattice distortions in a material to modify specific physicochemical properties. Strain engineering is applied for basic fundamental studies of physics and chemistry of solids but also for device fabrication through the development of materials with new functionalities. Thin films are one of the most important tools for strain engineering. Thin films can in fact develop large strain due to the crystalline constrains at the interface with the substrate and/or as the result of specific morphological features that can be selected by an appropriate tuning of the deposition parameters. Within this context, the in situ measurement of the substrate curvature is a powerful diagnostic tool allowing a real time monitoring of the stress state of the growing film.

This manuscript reviews a few recent applications of this technique and presents new measurements that point out the great potentials of the substrate curvature measurement in strain engineering. Our study also shows how, due to the high sensitivity of the technique, the correct interpretation of the results can be in certain cases not trivial and require complementary characterizations and an accurate knowledge of the physicochemical properties of the materials under investigation.

## I. Introduction

Many examples are known where a change of the interatomic distance in the lattice of a material leads to variations of certain physicochemical properties. Lattice strain can in fact modify conducting,[1-4] electronic and optical,[5] catalytic[6] or electrochemical,[7,8] mechanical and thermal[9] properties. The effect of strain can be in some case enabling new properties or functionalities that were not present in the relaxed structure. Many studies on the effect of strain are conducted using thin films as model systems where strain arises in consequence of the interfacial constrain at the film/substrate interface or due to specific morphological features that can often be selected by an appropriate tuning of the deposition parameters.

In the case of highly ordered epitaxial films, i.e. when film and substrate materials have similar lattice parameter and suitable crystallographic matching, strain is induced by the film-to-substrate lattice misfit. In the ideal case of a 1:1 match of all lattice planes at the interface, the lattice mismatch is entirely converted into lattice strain. Often however, a large part of the theoretical lattice misfit is compensated by introducing crystal defects to release the excess strain. A typical mechanism of stress relaxation is the formation and migration of misfit dislocations and for many oxides lattice strain exceeding a few percent cannot be elastically accomodated.[10]

While for epitaxial films with very high crystallographic quality the origin of the strain is obviously recognisable, for polycrystalline or textured films it is not straightforward. In this case in fact, the strain state and extent depend on the kind of grain boundary formed which in turn depends on the specific material, deposition method and experimental condition. Only the direct observation of the local morphological features, for example by transmission electron microscopy, can help to explain the measured strain state and identify its origin.

The quantitative analysis of strain in thin films is mostly performed ex situ, i.e. after the growth, by X-ray diffraction (XRD) typically through 2θ/θ scans and/or reciprocal space mapping. Of course these methods provide the overall and average value of strain along the film and cannot be used to investigate the evolution of the strain during the growth. This implies that the presence of regions with different strain at different distances from the substrate for example cannot be distinguished. Moreover, the analysis becomes very challenging in the case of ultra-thin layers (below 10 nm), which is often the range of thickness where large strain are retained. In situ XRD at synchrotron light sources[11] or reflection high energy electron diffraction (RHEED) were used as diagnostic tools to monitor the strain evolution. The first cannot obviously be routinely applied at laboratory scale on a daily basis. Concerning RHEED, its application as strain monitor is very rare.[12] The sensitivity of RHEED to the changes of the in-plane lattice parameter is limited by the spatial resolution of the diffracted electron spots. This is especially the case at relatively high pressure (often used in pulsed laser deposition or sputtering for example) and when the intensity of the spot changes with increasing thickness.

Optical measurements of the curvature of the substrate during the film growth may offer an efficient and practical tool to monitor in situ the direction and the evolution of the strain along the films.[13] The mechanic constraint of the substrate does not allow the film to grow freely along the plane of the surface of the substrate (in-plane). Volumetric changes without constraint in the film are only possible in the direction normal to the surface (out-of-plane). This results in stress generation along the growing film which exerts a force in-plane that bends the substrate. As the system reacts in order to minimize the elastic energy, the substrate develops a positive curvature (with respect to the substrate surface normal) in case of in plane tensile strain and negative curvature in case of compressive strain. The substrate curvature variation can be therefore used as a tool to identify the stress direction and evolution. Moreover, if the mechanical properties of the substrate are known also a quantitative estimation of the stress is possible.

Very accurate measurements of curvature variations can be obtained by measuring the deflection of a laser beam reflected from the bending substrate toward a CCD camera that records the changes of the position of the laser spot. For this purpose substrate in the form of thin, flexible cantilever were used.[14] The main limitation of such an approach is the availability of cantilevers made out of the desired material with the required crystallographic properties in terms of lattice mismatch and surface termination. So far mainly Pt and Si cantilever were used.

A more advanced experimental setup based on the same working principle is the so-called multi-beam optical stress sensors (MOSS). The MOSS uses an array of laser beams that are reflected from the surface of the substrate toward a CCD camera. Any change of the curvature of the substrate changes the relative distance among the laser beams at the CCD camera as schematically shown in Figure 1.

Such an experimental setup allows the use of any kind of substrates and using an n × m array of laser beams the variation of the distance between the spots can be measured accurately by averaging over multiple spots.

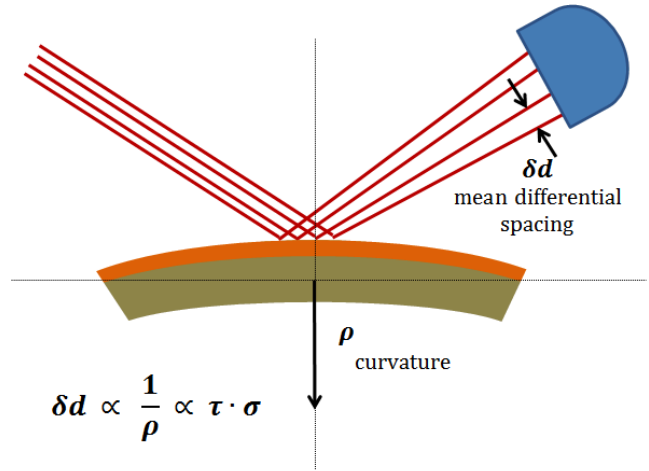

**Figure 1.** Schematic representation of the working principle of the Multi-beam Optical Stress Sensor. An n × m array of laser beams is reflected at the substrate surface toward a CCD camera. The mean differential spacing between the laser spots is recorded during the film deposition. This allows for monitoring the substrate curvature which is directly related to the stress.

The variation of the substrate curvature during the growth of the film can be measured by measuring with the MOSS the change of the mean differential spacing (m.d.s.) between the laser spots using the following equation:

$$\frac{1}{\rho} = -\frac{\cos\alpha}{2L}\frac{\delta d}{D_0}$$

where $\rho$ is the curvature, $\alpha$ is the angle of incidence of the laser beams with respect to the substrate surface normal, $L$ is the optical path of the laser beams, $D_o$ is the initial value of the m.d.s. and $\delta d$ is the variation of the m.d.s. as measured in situ by MOSS.

Once the changes of curvature can be measured, and knowing the elastic properties of the substrate, the Stoney equation can be used to calculate the stress thickness product (σ · τ) of the growing film:

$$\frac{1}{\rho} = \frac{6}{\tau_s^2} \frac{1-\nu}{Y} \sigma \cdot \tau$$

$\nu$, $Y$, and $\tau_s$ being respectively the Poisson ratio, Young modulus and thickness of the substrate. Literature reports several studies of stress generation and evolution in this films based on the MOSS analysis. Mainly metals and semiconductor films grown by ultra-high vacuum evaporation and sputtering[15-21] and pulsed laser deposition[22] (PLD) were investigated, and more recently also epitaxial oxides films made by PLD. [1,2,10] In situ investigation of the stress evolution in oxide films are indeed quite scarce, which is surprising considering the scientific and technological interest of oxide materials and the effect that strain can have in oxides.

In our previous studies[1,2] we used oxygen ion and proton conducting oxides as model systems to monitor real time by MOSS the evolution of stress. These solid state ionic conductors are materials of great scientific and technological interest for sustainable energy conversion.[23,24] The investigated films were epitaxially oriented and showed high crystallographic quality with only small-angle grain boundaries. Basically, surface energy and stress govern the initial stage of the growth, while the growth mode (layer-by-layer or island-like) and the nucleation and migration of dislocations lines determine the subsequent stress relaxation.[10]

With the present manuscript we report more details about the highest sensitivity we could achieve for the MOSS measurement of the substrate curvature during the growth of oxide materials by PLD (probably the most widespread method for growing oxide films) at high temperature in an oxygen background pressure. We also present new observations of the stress evolution in textured oxide films, with columnar polycrystalline morphology very commonly observed in oxide film deposited by PLD. Finally, we report on the importance of the choice of the substrate material for a meaningful interpretation of the in situ curvature measurements.

## II. Methods

Thin films of 15% Sm-doped $CeO_2$ (SDC) and 8% $Y_2O_3$ stabilized $ZrO_2$ (YSZ) were grown by pulsed laser deposition using sintered ceramic pellets prepared in our laboratories as targets for ablation. Commercially available 10×10×0.5 mm$^3$ single crystal of Al$_2$O$_3$, LaAlO$_3$, and SrTiO$_3$ were used as substrates. The vacuum chamber has a base pressure of about 10$^{-6}$ Pa and is equipped with a multi-beam optical stress sensor (MOSS) for in situ measurement of changes of the substrate curvature during the growth of thin films. For MOSS measurements a square 3×3 array of laser beams was directed to the centre of the substrate under an incidence angle of 30°. The distance between the spots of the laser beams in the array was in the range of 1 mm. O$_2$ was used as the background gas during ablation setting a partial pressure in the range between 2 and 5 Pa. The target to substrate distance was set at 5 cm. A radiant heater was used to set the deposition temperature at 750 °C. The measurements of the changes of the curvature of the substrates were always performed at constant temperature and background pressure. The MOSS curvature measurement does not allow the use of a metal paste to provide the required thermal contact between the substrate and heating stage. For this, the back (unpolished) side of the substrates was coated by a sputtered Pt film, about 500 nm thick, acting as the heat absorber for the otherwise transparent substrates. The deposition temperature was read out using a pyrometer pointing at the centre of the substrate setting the emissivity value of 0.97 for black Pt. A 248 nm KrF excimer laser with pulse width of 25 ns was focused onto the targets on a spot of about 1 mm$^2$ with an energy density of about 1.3 J cm$^{-2}$. In these experimental conditions, with a laser frequency of 2 Hz, a deposition rate of about 0.1 and 0.07 Å per pulse was found for SDC and YSZ, respectively. The deposition rate was calibrated by X-ray reflectometry. The structural characterization of the films was performed by X-ray diffraction.

### III. Results and Discussion

We focus first on the sensitivity that could be achieved with our experimental setup for the measurement of the relative change of the curvature of the substrate during the growth.
We make use of a thin film of 8% Y$_2$O$_3$ stabilized ZrO$_2$ (YSZ) grown on Al$_2$O$_3$ substrate as an example. Figure 2a shows the XRD analysis of the film revealing the polycrystalline nature with a textured microstructure characterized by grains (100) and (111) out-of-plane oriented.
Figure 2b shows the X-ray reflectometry (XRR) measurement used for the calibration of the deposition rate which was found to be 0.07 Å per pulse with the selected deposition parameters. The red open circles in Figure 2c draw the progress of the substrate curvature with time. The curvature is calculated by averaging the m.d.s. measured by MOSS. The continuous black line shows how the thickness of the film changes with time. The slope of the curve is calculated on the base of the XRR measurement of Figure 2b.

The deposition starts at 1000 seconds and the MOSS shows a negative curvature of the substrate which indicates the development of an in-plane compressive stress along the growing film (Figure 2c). The curvature increases almost linearly with increasing thickness. According to the Stoney equation this indicates an almost constant stress value.

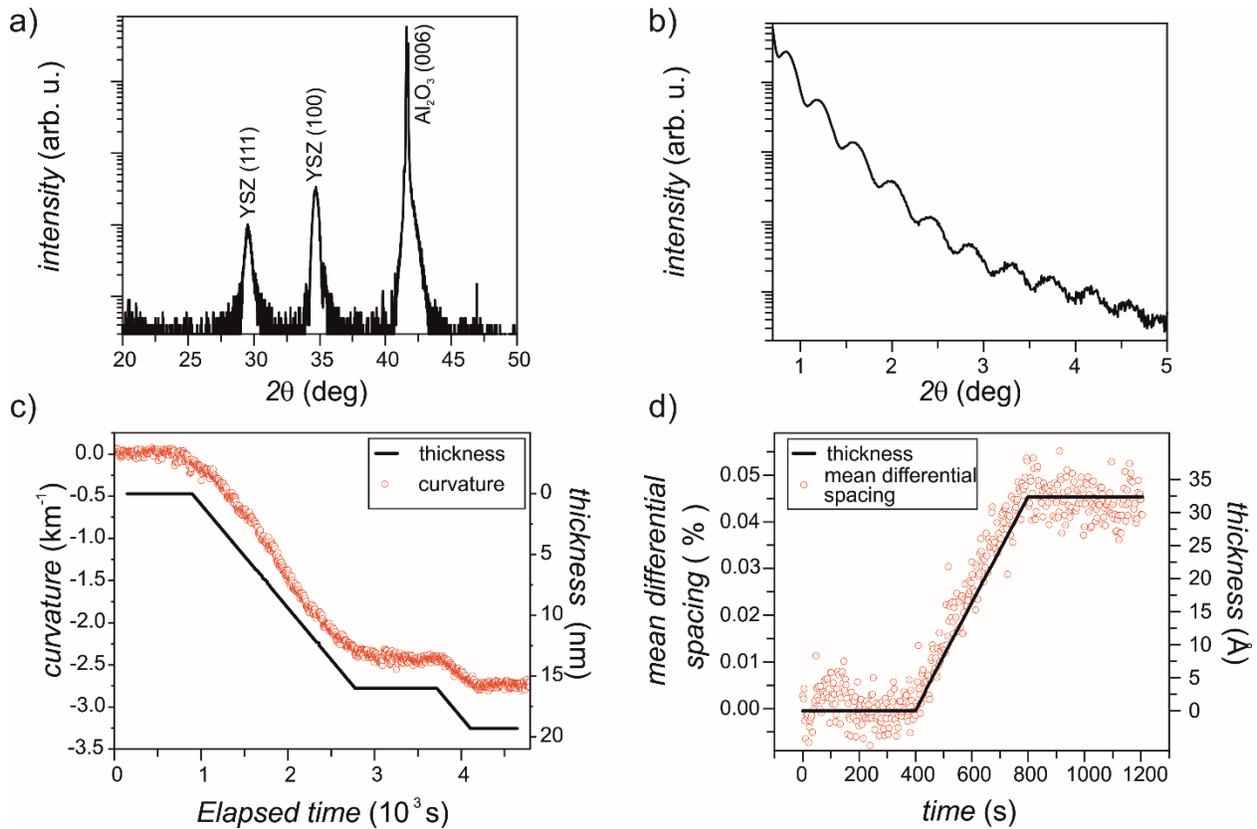

**Figure 2.** Analysis ex situ of the crystal structure and in situ of the stress evolution in a thin film of YSZ on $Al_2O_3$ (0001) a) 2θ/θ scan of the polycrystalline thin film. Film peaks correspond to the (100) and (111) orientations. b) XRR scan used for thickness measurement and deposition rate calibration of YSZ thin films. c) Red circle: Substrate curvature variation monitored by MOSS as a function of time. The negative substrate curvature indicates in-plane compressive strain. Black line: film thickness calculated from the XRR measurement of the deposition rate. Thin film deposition occurs between 1000 and 2800 s and between 3600 and 4000 s. Variation in the substrate curvature are already distinguishable at 8 – 10 Å thickness. d) Red circle: Mean differential spacing of the laser beam spots recorded by MOSS (MOSS raw data) during the deposition of the topmost 3 nm of the same sample. Deposition occurs between 400 and 800 s. Black line: Calculated thickness as a function of the time.

The deposition is stopped after 2800 seconds and the MOSS shows that also the curvature stops increasing. The elastic energy accumulated in the film in the form of a compressive in-plane strain stays constant when no more material is added. It is important to note that no evidence of stress relaxation can be observed. At this point the thickness of the YSZ film is about 17 nm. After 1000 seconds the deposition was resumed and the MOSS clearly shows that the sapphire substrate continues bending in the same direction. After the growth of additional 3 nm the deposition is ended with a total thickness of about 20 nm and the MOSS shows again that no further changes of the substrate curvature are detected. In Figure 2d the open red circles indicates the in situ MOSS measurement of the m.d.s. (the MOSS raw data) as a function of time and the black line is the evolution of the thickness of the film. This data refers to the last 3 nm added after the first stop of the deposition process shown in Figure 2c. We would like to highlight the high sensitivity of the technique, which depends on the background noise, the substrate elastic modulus and the stress generated in the film during the deposition.

For $Al_2O_3$ substrate and the observed signal-to-noise ratio (depending on the substrate material and the background vibrations) the curvature resolution leads to a detectable variation of the strain-thickness product of -2.5 GPa nm already at thickness 1 nm; which means that for strains close to 0.4% and an elastic modulus of 350 GPa (as in the case of $Al_2O_3$), clear changes of wafer curvature can be detected already for thicknesses as small as of 8-10 Å (corresponding to less than 2 unit cells of YSZ). We would like also to highlight here that MOSS analysis not only provides a real-time diagnostic of the stress state of the growing films (whether compressive or tensile) but it provides reliable information for film thicknesses that would be almost impossible to analyse by standard (ex situ) XRD.

The use of C-cut sapphire substrates can promote the epitaxial growth of YSZ films along the (111) crystallographic direction.[25] Also the textured morphology with mixed (100) and (111) orientation is reported.[26] In general, as for many oxides, films prepared by PLD show a typical columnar morphology consisting on parallel pillars with relatively narrow size distribution separated by grain boundary regions.[25,27] An example of such morphological feature is given in Figure 3 for a film of doped ceria grown on a sapphire substrate.

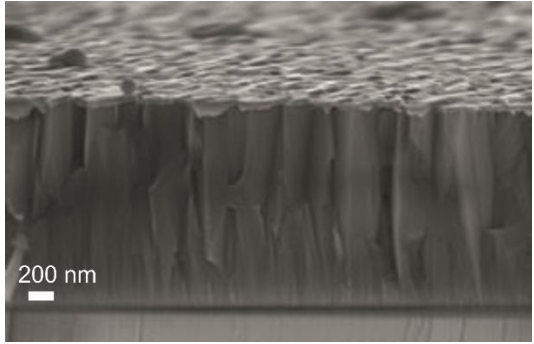

**Figure 3**: SEM micrograph of a thin film of doped ceria on $Al_2O_3$ (0001). The typical columnar morphology of the film is clearly visible.

In the case of epitaxial films of YSZ on C-cut sapphire, due to the lattice mismatch an in-plane compressive strain of the film is expected, which is the same strain state detected by MOSS for our samples. The out-of-plane lattice parameter calculated from the XRD measurement was about 5.16 Å indicating an out-of-plane tensile strain consistently with the in-plane compressive strain observed by MOSS. However, due to the columnar morphology, even in the case of epitaxial films it would be questionable to ascribe a measured in-plane compressive strain to the lattice misfit. Complementary characterizations (transmission electron microscopy, for example) would be needed to investigate what type of grain boundary is present between adjacent grains and how the grain boundary regions contribute in determining the final stress of the film. This consideration is even more important in the case of textured films showing multiple orientations, as those reported here.

C-cut sapphire substrates were also used for the growth of SDC films. Epitaxial films[28] as well as films showing multiple orientations are reported in the literature. The large lattice mismatch can favor the formation of interfacial misfit dislocations leading to the growth of relaxed epitaxial films with almost no evidence of grain separation.[28] Instead, when polycrystalline films were grown with the typical columnar morphology using $SiO_2$ substrates, the strain was found to be compressive in-plane (tensile out-of-plane).[28]

Figure 4a shows the X-ray diffraction pattern of a thin film of SDC grown on (0001)-oriented sapphire substrate. The film is polycrystalline with multiple orientations. Figure 4b shows the MOSS measurement of the substrate curvature (red open circles) during the growth and the evolution with time of the film thickness (black line), as calibrated by XRR. In agreement with literature,[28] for polycrystalline films the in-plane stress is compressive, as can be observed real-time by MOSS (negative substrate curvature). Under the selected deposition parameters, up to a maximum thickness of about 20 - 25 nm the curvature remains almost constant. The same behavior was observed for YSZ films, as can be seen in Figure 2c. Growing the thickness larger (above 25 nm), the slope of the curve in Figure 4b decreases indicating that the stress is released. The total thickness of this film is about 35 nm but the topmost 10 nm do not contribute to the overall strain. The XRD analysis reveals an average out-of-plane tensile strain of about 0.15% which confirms the MOSS observation of an in-plane compressive stress. The comparison of these measurements with the structural analysis reported in reference [28] suggests that for both ceria and zirconia the typical columnar morphology obtained by PLD leads to the development of an in-plane compressive stress.

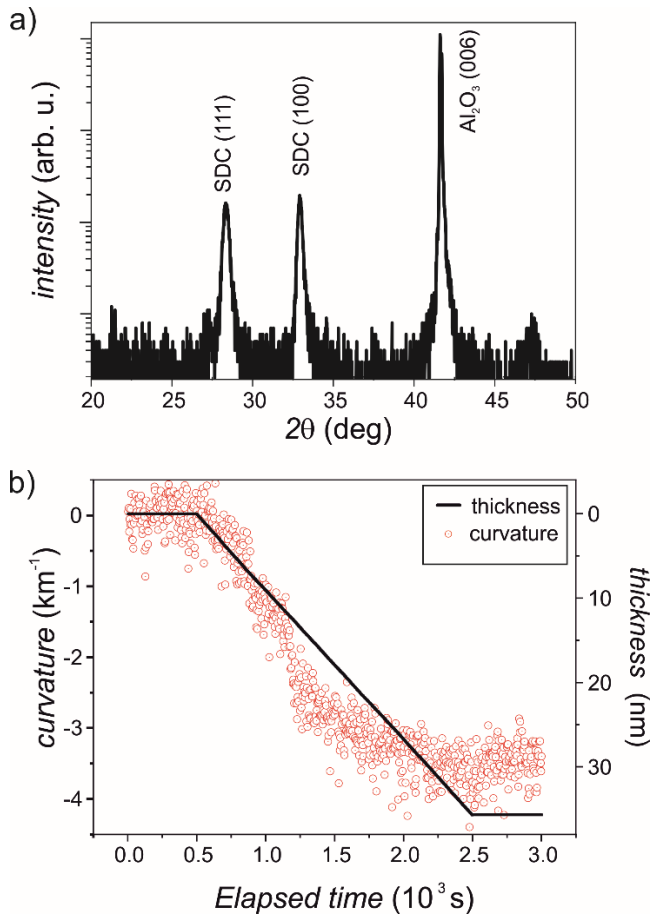

**Figure 4.** a) 2θ/θ scan of a polycrystalline thin film of SDC on $Al_2O_3$ (0001) substrate. Film peaks correspond to the (100) and (111) orientations. b) Substrate curvature (red circles) recorded by MOSS during the deposition of the same sample. Black line indicates the films

thickness calculated according to the deposition rate. Thin film deposition occurs between 500 and 2500 s. The variation in the curvature line slope around 1000 s indicates that the stress is released.

Finally, we would like to report a remarkable effect observed using SrTiO$_3$ (STO) as substrate. STO is among the most commonly used substrates for thin film growth; however the importance of the stability of its chemical composition, in term of oxygen content, is often underestimated. STO, as other cubic or pseudocubic perovskite substrates such as LaAlO$_3$ (LAO) or NdGaO$_3$ (NGO), is frequently used for the growth of highly ordered epitaxial ceria films.[1] STO was also used as buffer layer deposited on MgO substrates for the growth of ceria films [29,30] or ceria/zirconia multilayers.[30-32]

The potential problem of this material is that at high temperature and relatively low oxygen partial pressure (i.e. in the typical condition for thin film growth by PLD) STO is easily reduced creating oxygen vacancies and thus enabling ionic and electronic conductivities. The ionic mobility allows oxygen ions to diffuse easily across the interface with the growing film. Using $^{18}$O-labelled STO substrates it was shown that the STO substrate itself can become the main source of oxygen for the growing film, even more than the oxygen molecules in the surrounding gaseous environment of the target material when low background pressure is used.[33,34] This may have very important consequences as far as the stress generation and evolution in the growing film is concerned, as described in Figure 5.

Figure 5a shows the XRD analysis of two films of approximately the same thickness (36 – 38 nm) of SDC grown on LAO (100) and STO (100). In both cases we have films (100)-oriented. The inset in Figure 5a shows the magnification of the angular region around the (200) reflexes. The dashed lines indicates the angular position of the (200) diffraction peak of the relaxed structure of 15% Sm-doped CeO$_2$ with a lattice parameter of about 5.43 Å. [35-37]

As can be seen, compared to the relaxed structure both films show an out-of-plane tensile strain which is about 0.40% for the film grown on LAO and 0.15% for that grown on STO. This implies the presence of an in-plane compressive strain of SDC which is in agreement with what one would expect on LAO (with a lattice mismatch of about 0.49%) but it does not agree with what is expected on STO. On STO in fact, the SDC film should develop an in-plane tensile strain as the consequence of a lattice mismatch of about 1.6%.

The MOSS measurements anticipated in situ the conclusion obtained ex situ by XRD, as shown in Figure 5b. In both cases we measured a positive change of the m.d.s. between the laser spots of the MOSS, indicating the development of an in-plane compressive stress.

On the LAO substrate the stress rises at the very early stage of the growth. The slope of the curve describing the evolution of the m.d.s. vs. time decreases slightly after 1000 seconds (corresponding to a film thickness of 25 nm) indicating that part of the strain is released.

Instead, MOSS shows a very different stress evolution during the SDC deposition on the STO substrate. Negligible curvature variation of the substrate was detected for the first 10 - 15 nm.

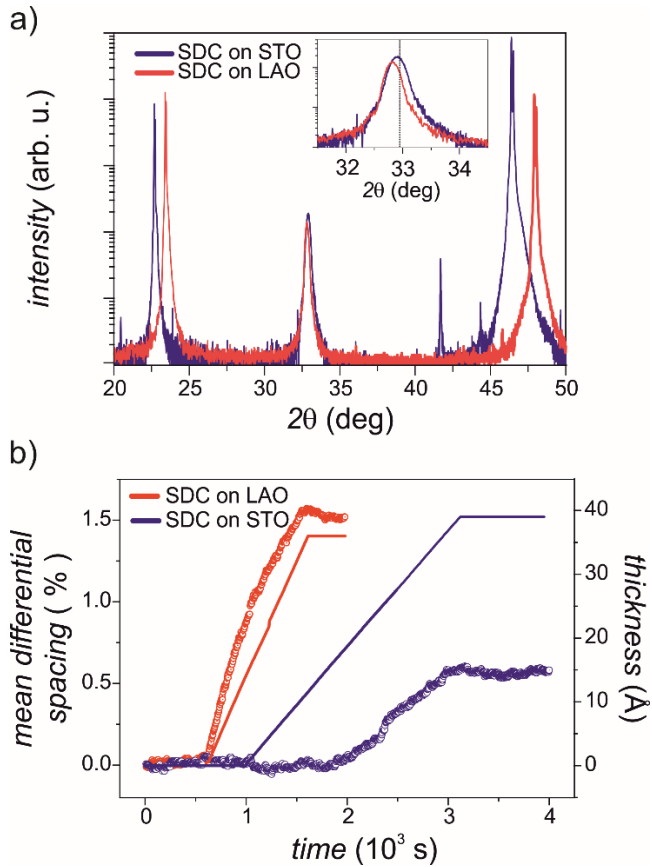

**Figure 5.** a) XRD diffraction pattern of thin films of SDC on STO and LAO. In the inset: magnification of the (200) reflex. b) Circles: MOSS analysis of the SDC layers during growth on STO and LAO substrates. Lines: thickness increment as a function of time. Line slope was determined according to the deposition rate.

After that, the m.d.s. starts increasing with almost constant slope. The m.d.s. remains then constant when the deposition ends. The same qualitative behaviour was observed for several SDC films on STO (always in-plane compressive stress was measured), though the details of the curvature vs. time curve can vary quite significantly from sample to sample. On the contrary, the MOSS measurements were very well reproducible using LAO or NGO substrates,[1] materials that are more difficult to reduce. It is also interesting to note that on LAO and NGO, not only the stress state of SDC is in line with the lattice misfit (in-plane compressive on LAO and tensile on NGO), but also the strain measured ex situ by XRD showed values very similar to the lattice misfit.[1]

Summarizing, using STO as a substrate the in situ MOSS measurements are less reproducible and, in agreement with ex situ XRD, show a stress state of the film opposite to that expected considering the lattice misfit. High quality ceria films are typically obtained with deposition parameters similar to those used in the present study using cubic or pseudo-cubic perovskite substrates such as NGO, LAO and STO. The compressive, instead of tensile, stress observed for the SDC films on STO is not ascribed to the effect of the textured polycrystalline morphology, as in the cases of the ceria and zirconia films grown on $Al_2O_3$ discussed above.

A possible way to rationalize the experimental observation is the aforementioned high oxygen ion mobility in STO and its tendency to be reduced when exposed to the typical deposition conditions of high temperature and relatively low oxygen background pressure. As observed in [33] and [34] for films of LAO and YSZ, the STO substrate can become the main source of oxygen for the growing film which act as a sort of oxygen pump for the substrate. Due to the fast oxygen ion diffusion in SDC, the oxygen ion concentration is expected to equilibrate through the film thickness (40 nm) quite fast and no gradient of the oxygen ion concentration is expected in the film after the deposition, as observed for YSZ.[34] Conversely, oxygen ion will be depleted from an STO layer at the film/substrate interface and in this interfacial layer the more reduced STO will have a slightly larger lattice parameter than the less reduce STO in the bulk. Within this scenario, the formation of such a layer would have the equivalent effect of the deposition of thin film of a material in compressive in-plane stress that would bend the substrate inducing a positive variation of the m.d.s. of the laser beams of the MOSS (equivalent to negative substrate curvature variation), as shown in Figure 5b. This mechanism could thus explain the sign of the curvature variation of the substrate detected by MOSS, and may be the cause of the final in-plane compressive stress of the SDC films. The mechanisms through which the tensile strain in SDC is released cannot be directly identified and can vary for different deposition parameters and therefore different lattice parameter of STO at the interface. Wafer curvature was in fact reported to be a reliable way to study the variation of oxygen concentration in metal oxides.[38]

Higher oxygen background partial pressure, in the range of a few tens of Pa, may be enough to keep the oxygen content of the substrate more stable. Also, the use of $O_3$ or $N_2O$ as the background gas may increase the amount of oxygen in the film originating from the gaseous environment, thus reducing the oxygen ion exchange with the substrate.

The measurements reported here show that only materials with highly stable oxygen content are recommended as the substrate for application in oxide strain engineering. Very common materials such as STO may lead to unpredictable results.

## IV. Conclusions

In situ wafer curvature measurements offer a powerful tool for strain engineering to monitor real-time the evolution of stress along a growing film. The application of this diagnostic method for oxide materials is still quite rare, though a remarkable sensitivity on the stress state of the film can be achieved. We have shown here that changes of substrate curvature induced by the deposition of less than 2 unit cells of the growing film can be clearly detected.

The multi-beam optical stress sensor provides invaluable qualitative insights into the evolution of the stress in-plane and in situ to complement the quantitative measurement of the out-of-plane strain performed ex situ by standard $2\theta/\theta$ scan.

Wafer curvature measurements are highly reliable and reproducible when stable oxide materials are used as the substrates. This is the case for example for $Al_2O_3$, MgO, $LaAlO_3$, $NdGaO_3$. Special care has to be taken when the chemical composition of the substrate is not stable, as it is the case for instance for $SrTiO_3$, one of the most commonly used single crystal substrate also for application in strain engineering.

## Acknowledgements

The authors gratefully thank the Swiss National Science Foundation under grant agreement number v200020_147190.